# Two-dimensional alternating ferrimagnetism with strain-controlled half-metallic state and valley polarization

W. Z. Zhuo[1*], Z. H. Guan[2], Z. L. Peng[2], Y. N. Pan[3], J. Chen[4*], Y. Yang[3*], and M. H. Qin[2*]

[1] *School of Optoelectronic Engineering, Guangdong Polytechnic Normal University, Guangzhou 510665, China*

[2] *Guangdong Provincial Key Laboratory of Quantum Engineering and Quantum Materials and Institute for Advanced Materials, South China Academy of Advanced Optoelectronics, South China Normal University, Guangzhou 510006, China*

[3] *School of Physics, Hefei University of Technology, Hefei 230009, China*

[4] *Key Laboratory of Quantum Materials and Devices of Ministry of Education, School of Physics, Southeast University, Nanjing 211189*

The discovery of altermagnetism offers new opportunities for exploring novel quantum states and developing spintronic devices for enabling momentum-dependent spin splitting in compensated systems, while zero net magnetization limit its manipulability using conventional magnetic method. Here, we propose 2D alternating ferrimagnetism — a phase merging alternating momentum-dependent spin-splitting with a finite net magnetization. A tight-binding model reveals that alternating ferrimagnetism originates from uncompensated magnetization in altermagnets, facilitating concurrent net magnetization and alternating spin-splitting. First-principles calculations and Monte Carlo simulations demonstrate stable alternating ferrimagnetism in strained and Cr-substiting $V_2Te_2O$, which exhibit strain-tunable net magnetization, reversable half-metallicity and valley polarization, accompanied by long-range magnetic order above room temperature. By combining altermagnetic and ferromagnetic properties, alternating ferrimagnetism expand the 2D magnetism landscape and offer pathways for energy-efficient spintronic applications.

## 1. Introduction

Collinear magnetic orders have always played important roles in spintronic devices such as magnetic memory and logic devices. [1-2] For example, ferromagnets (FMs) in which all magnetic moments are parallelly aligned dominate almost the whole data storage market. [3-9] In contrast to ferromagnets, antiferromagnets (AFMs) hold antiparallel arranged magnetic moments and great potential in high density and ultrafast device design due to their zero net magnetization and ultrafast dynamics. [10,11] In conventional AFMs, however, compensated magnetic sublattices connected by the symmetry inversion $P$ or translation $\tau$ operations lead to spin-degenerate electronic bands, limiting their utilization in spin-dependent devices. [10,11] Interestingly, the recently proposed altermagnets (AMs) exhibit momentum-dependent spin splitting even in the absence of spin-orbit coupling, owing to their specific crystal symmetries connecting opposite-spin sublattices, which quickly draw significant attentions in spintronics and materials science. [12-14] So far, altermagnetism in bulk systems like MnTe, [15,16] $RuO_2$, [17,18] CrSb [19,20] and $KV_2Se_2O$ [21] have been experimentally observed, contributing to the rapid development of altermagnets, while two-dimensional (2D) altermagnets are mainly limited to theoretical explorations which also demonstrates their promising novel physical property. [22-30]

Generally, zero net magnetization in conventional antiferromagnets and altermagnets precludes modulation of the Néel vector by conventional magnetic methods, hindering their practical applications. [31,32] Moreover, band splitting in ferromagnets is nearly independent on momentum, which strongly limits the polarization efficiency of spin current. [33] In some extent, a viable strategy to circumvent these issues involves leveraging the coupling between ferromagnetisms and altermagnetisms. To be specific, introducing spin disparity between symmetry-connected sublattices in 2D altermagnets may induce a novel phase that concurrently exhibits altermagnetic momentum-dependent spin splitting and nonzero net magnetization. This dual properites render the system, naming alternating ferrimagnets (AFiMs), simultaneously manipulable of the Néel vector through conventional magnetic methods and capable of generating non-relativistic momentum-dependent spin currents. Furthermore, the synergistic combination of alternating spin-splitting bands and spin-dependent band shifting in AFiMs may give rise to exotic electronic properties including half-metallicity [34,35] and valley polarization. [36,37] Thus, AFiMs may generate a fully polarized spin current and exhibit valley Hall effect, allowing AFiMs the promising material platform for next-generation spintronic devices. Thus, a systematic study on alternating ferrimagnetism urgently deserves to be performed to provide information for future experiments and applications.

In this work, we propose and theoretically verify the existence of 2D alternating ferrimagnets. Using a tight-binding model, we theoretically demonstrate that introducing disparity of local magnetic moments at two sublattices generates a finite net magnetization, while the alternating spin-splitting is remained. Two feasible routes to achieve 2D alternating ferrimagnets are proposed: strain engineering and elemental substitution in altermagnetic monolayers such as $V_2Te_2O$. Our first-principles calculations and Monte Carlo simulations reveal strain-tunable band structures, half-metallic behavior and valley polarization with long-range magnetic order above room temperature. These behaviors are supported by calculated electronic structures, Fermi surfaces, and finite-temperature magnetic properties. This study unveils the alternating ferrimagnetism through combining the altermagnetism and ferrimagnetism, opening avenues for designing multifunctional spintronic devices with enhanced performance and integrability.

## 2. Result

In this work, we study a 2D tight-binding model [26,27] to describe ferromagnets, antiferromagnets, altermagnets, and alternating ferrimagnets, as well as phase transitions between these magnetic states. This model is constructed based on symmetry considerations, and the model Hamiltonian is defined as:

$$\begin{aligned} H = & t\sum_{i,\delta_1}(c^+_{a,i}c_{b,i+\delta_1} + h.c.) + \sum_i (t^a_x c^+_{a,i}c_{a,i+\delta_{2x}} + t^a_y c^+_{a,i}c_{a,i+\delta_{2y}} + t^b_x c^+_{b,i}c_{b,i+\delta_{2x}} + t^b_y c^+_{b,i}c_{b,i+\delta_{2y}} + h.c.) \\ & -\frac{\Delta}{2}\sum_i c^+_{a,i}c_{a,i} + \frac{\Delta}{2}\sum_i c^+_{b,i}c_{b,i} + \sum_i M_a c^+_{a,i}c_{a,i}s_z + \sum_i M_b c^+_{b,i}c_{b,i}s_z \end{aligned} \tag{1}$$

Where $t$ represents the nearest-neighbor hopping integral, and $t_x^a$, $t_x^b$, $t_y^a$ and $t_y^b$ are the next-nearest-neighbor hopping parameters, as illustrated in Figure 1(a). $\Delta$ is the staggered potential, $M_a$ and $M_b$ are the magnetizations of two sublattices, and $s_z$ is the spin operator.

$\delta_1$ and $\delta_{2x/2y}$ denote the nearest and next-nearest vectors, respectively. These parameters can be used to regulate the symmetry of the system, corresponding to the cases of FMs, AFMs, AMs, and AFiMs.

For $t_x^a = t_x^b = t_y^a = t_y^b$, $M_a = M_b$, and $\Delta = 0$, the system shows a conventional spin-dependent band shifting due to the broken time reversion symmetry *T* in ferromagnets (Figure S1(a)). When the parameters satisfy $t_x^a = t_x^b$, $t_y^a = t_y^b$, $M_a = -M_b$, and $\Delta = 0$, the system exhibits inversion symmetry *P*, which connects the spin-opposite sublattices. In this case, a combined *TP* symmetry exists with the time-reversal operator *T* reversing spins, leading to spin-degenerate bands characteristic of conventional AFMs, as illustrated in Figure S1(b). In contrast, the conditions $t_x^a = t_y^b$, $t_y^a = t_x^b$, $M_a = -M_b$, $\Delta = 0$, and $t_x^a \neq t_x^b$ break *P* and $\tau$ symmetry but preserve a $C_4$ rotation symmetry, swapping $t_x^a$ with $t_y^b$, and $t_y^a$ with $t_x^b$, yielding alternating momentum-dependent spin splitting, as illustrated in Figure 1(b).

We now consider ferrimagnetic condition characterized by a disparity in the magnetization magnitudes between two sublattices (i.e., $|M_a| \neq |M_b|$), which necessitates an introduction of a nonzero staggered potential ($\Delta \neq 0$). It is evident that this configuration breaks both the *PT* symmetry and the $\{T|C_4\}$ symmetry. However, when the hopping parameters satisfy $t_x^a \approx t_y^b$ and $t_y^a \approx t_x^b$, the system not only retains the alternating spin-splitting of altermagnets but also exhibits a spin-dependent band shifting concomitant with a finite net magnetization, as depicted in Figures 1(c)-1(f). For $|M_a| \neq |M_b|$ and a nonzero $\Delta$, a spin-dependent band shifting similar to that in ferromagnets emerges, and the alternating spin-splitting feature preserves, as illustrated in Figure 1(c). Notably, the energy bands remain gapless in this regime because of small disparity in magnetization magnitudes ($|M_b| - |M_a|$) and weak $\Delta$. Upon increasing $|M_b| - |M_a|$ and $\Delta$, a band gap emerges for one spin channel, and the other channel remains nearly gapless (Figure 1(d)).

Interestingly, as the hopping parameters decrease, a half-metallic state is achieved, wherein the Fermi surface intersects the spin-up band and crosses the gap of spin-down channel, as depicted in Figure 1(e). Furthermore, polarized valleys located at X and Y points are induced, as shown in Figure 1(f). The half-metallic state and valley polarization effect are attributed to the concurrent alternating band splitting and spin-dependent band shifting in AFiMs through tuning these hopping parameters. Besides, the Néel vector can be controlled through conventional magnetic method due to nonzero net magnetization. Thus, AFiM provide a novel platform to realize and control half-metallicity and valley polarization which are the cornerstones of next-generation spintronic and valleytronic devices. [34-37]

The proposed 2D alternating ferrimagnetism exhibit concurrent alternating band splitting and spin-dependent band shifting, establishing a novel pathway for exploring exotic quantum states. For its realization in practical materials, introducing spin disparity between two sublattices connected by specific crystal symmetries in 2D altermagnets is one promising way. Here, we propose two feasible approaches: strain engineering and elemental substitution.

As a representative candidate, altermagnetic monolayer $V_2Te_2O$ with tetragonal structure is investigated, wherein magnetic vanadium atoms are coordinated by nonmagnetic tellurium and oxygen ligands, as depicted in Figures 2(a) and 2(b). The calculated in-plane lattice constant ($a$ = 3.93 Å) is consistent with experimental report [39] (Table 1). The calculated phonon dispersion reveals the absence of imaginary frequencies (Figure S2(a)), confirming its dynamic stability. The altermagnetic ground state (labeled in Figure 2(b)) is confirmed through energy comparisons of various spin configurations (see Figure S3 and Table 2), in agreement with prior studies. [24] As shown in Figure 2(c), the electronic band structure of $V_2Te_2O$ exhibits distinct alternating momentum-dependent spin splitting. In particular, the band crossings around X and Y high-symmetry points forming separated topological nodal loops with opposite spin polarizations. The spin polarization is fully compensated between counterpart states around X and Y points on the spin-polarized Fermi surfaces, as depicted in Figure 2(d).

As predicted in the model analysis, uniaxial strain induces two inequivalent V atoms (Figure 2(e)) with a spin disparity, and thereby realizing an alternating ferrimagnetic state. As shown in Figure 2(f), under 5% tensile strain, a spin-dependent band shifting is induced with an upward shift of the spin-down band at X point and a downward shift of the spin-up band at Y point, while the alternating momentum-dependent spin-splitting still preserves. The uncompensated Fermi surfaces shown in Figure 2(g) further confirm the spin-dependent band shifting. Figure S4 presents the detail evolution of band structure in monolayered $V_2Te_2O$ with various uniaxial strains ranging from -8% to 8%, which demonstrates that both compressed and tensile strains can induce the spin-dependent band shifting with preserved alternating momentum-dependent spin-splitting.

Figure 2(h) summarizes the spin-dependent band shifting energy between X and Y points near the fermi level ($\Delta E = E(X) - E(Y)$, as labeled in Figure 2(f)). It is shown that both spin-dependent band shifts of valance and conduction bands vary monotonously with uniaxial strain, demonstrating a strain-controllable band shifting. This band polarization enables a nonzero net magnetization which can be estimated by the integration of the spin density within the energy range from negative infinity to Fermi level ($M_{\mathrm{net}} = \int_{-\infty}^{E_f} [\rho^{\uparrow}(\varepsilon) - \rho^{\downarrow}(\varepsilon)] \mathrm{d}E$). In unstrained case, the net magnetization vanishes due to full compensation. A spin-dependent band shifting is induced by applied strain, and the disparity in spin densities amplifies, resulting in a monotonically tuned net magnetization, as shown in Figure 2(i). It is worth noting that in 2D Janus $V_2SeTeO$, [25] the net magnetization cannot be induced by strain without ion doping due to its semiconducting property, which is different from 2D $V_2Se_2O$ investigated in this study. Moreover, the net magnetization in 2D $V_2Se_2O$ can be reversed via strain switching, highlighting its potential in spintronic applications. [31,32]

One notes that the strain-induced spin disparity of two sublattices is rather weak, as illustrated in Figure 1(c), while a pronounced polarization could be achieved through elemental substitution, according to our TB model analysis. To check this phenomenon, we investigate the effect of half substitution of V by Cr. Figure 3(a) depicts the structure of 2D $VCrTe_2O$, which also lifts the $C_4$ symmetry. The phonon dispersion calculation reveals the absence of imaginary frequencies (Figure S2(b)), confirming its dynamic stability. Furthermore, the altermagnetic ground state is also demonstrated by the energy comparisons presented in Table 2. Both the alternating momentum-dependent spin-splitting and spin-dependent band shifting are clearly manifested in the electronic structure (Figure 3(b)) and the spin-resolved Fermi surfaces of $VCrTe_2O$ (Figure 3(c)). The computed magnetic moments of V and Cr atoms are $-1.873$ $\mu_B$ and 3.046 $\mu_B$, respectively (Table 1), which generate a net magnetization and staggered potential significantly larger than that in strained $V_2Te_2O$, resulting in a gap near the Fermi level at Y point. However, the Fermi surface still intersects both spin-down and spin-up bands, prohibiting a half-metallic state. Interestingly, this deficiency can be compensated by applying biaxial tensile strain to decrease the hopping parameter while keeping the symmetry, as revealed below.

The calculated band structure shown in Figure 4 clearly illustrates the biaxial strain-dependent band splitting in $VCrTe_2O$. Under a 1% biaxial tensile strain, a band gap totally opens in the spin-up channel near the Fermi level, while the spin-down channel preserves a gapless nodal loop (Figure 4(a)), demonstrating the half-metallic state, consistent with the model prediction in Figure 1(e). The corresponding spin-polarized Fermi surfaces exhibit a single pocket near X point for the spin-down component (inset of Figure 4(a)). The half-metallic phase, accompanied by alternating momentum-dependent spin-splitting, may host intriguing spin Hall and anomalous Hall effects. [12-14,24] The half-metallicity preserves in a tensile strain range of (1% ~ 3%), as shown in Figure 4(b), benefiting future experimental exploration. As the tensile strain increases to ~ 6%, the two spin-down bands form a crossing point near the Fermi level along the Γ-X path. Surprisingly, under 7% tensile strain, an additional half-metallic state emerges, wherein the Fermi surface contrarily intersects the spin-up band, as further validated by the spin-polarized Fermi surfaces (inset of Figure 4(e)).

The strain-mediated transition between the half-metallic states with opposing spin polarizations near the Fermi level (under 1% and 7% tensile strains) in 2D $VCrTe_2O$ can be applied to generate fully polarized spin current, wherein the polarization direction of spin current can be manipulated by strain without an external magnetic field, which provides new avenues for spintronic applications. [1,3,34,35] In addition, the transition from the nodal-loop phase to the gapped phase under tensile strain implies a critical reconstruction of the band-crossing manifold. At the critical tensile strain of 6%, the original line-node degeneracy collapses, signaling a topological phase transition. [39,40] With further increasing the tensile strain to 8%, the system transits to a semiconducting state, as shown in Figure 4(f), primarily due to the progressively weakening of the hopping integrals induced by tensile deformation. Moreover, a valley polarization effect between X point and Y point is observed under strains above 7 %, which may give rise to valley Hall effect, [36,37,41] thus providing a novel platform for valleytronic devices.

The magnetic properties of $V_2Te_2O$ and $VCrTe_2O$ monolayers are systematically investigated. $V_2Te_2O$ exhibits perpendicular magnetic anisotropy, whereas $VCrTe_2O$ shows an easy axis aligned along the *b*-direction, as depicted in Figure S5. The out-of-plane anisotropy in $V_2Te_2O$ contributes to the long-range magnetic order in 2D limit,[42] and the in-plane anisotropy in $VCrTe_2O$ also allows long-range order due to the broken continuous symmetry. [42]

Finally, we explore finite-temperature magnetic behavior using Monte Carlo simulations based on the model Hamiltonian:

$$H = J_1 \sum_{<i,j>} \mathbf{S_i S_j} + \sum_{<<i,j>>} J_{2\alpha} \mathbf{S_i S_j} - K_x \sum_i (S_i^x)^2 - K_y \sum_i (S_i^y)^2 - K_z \sum_i (S_i^z)^2 \qquad (2)$$

Here, $\mathbf{S_i}$ denotes the spin vector at the $i$-th magnetic site (contributed by V/Cr atoms), $< i,j >$ and $\ll i,j \gg$ represent the first- and second-nearest-neighbor pairs, respectively. The Heisenberg exchange parameters $J_1$ and $J_{2\alpha}$ are determined by comparing energies of different spin configurations [24,30] (Table 2, Figure S3), while the anisotropy constants $K_x$, $K_y$ and $K_z$ are derived from energy calculations with spins aligned along the *a*, *b*, and *c* axes, [24,30] respectively. The optimized parameters are summarized in Table 3.

The temperature-dependent magnetization and Néel vector are presented in Figure 5(a) and (b), respectively, which demonstrate that the pristine 2D $V_2Te_2O$ maintains zero net magnetization at low temperatures, whereas $VCrTe_2O$ sustains a finite magnetization due to its ferrimagnetic ground state, confirming the ground state of the altermagneitc configuration. The estimated magnetic transition temperatures are approximately 650 K for $V_2Te_2O$ and 490 K for $VCrTe_2O$, well exceeding room temperature, which underscores their potential for practical applications.

## 3. Discussion

It is well noted that switching Néel vector in conventional antiferromagnets and altermagnets keeps a strong challenge, limiting their future applications. [31,32] In this work, the alternating ferrimagnetism is demonstrated through TB model analysis and first-principles calculations, which may provide an alternative way to avoid this challenge. In strained 2D $V_2Te_2O$, the spin-dependent band shifting generates net magnetization, and tensile strain and compressed strain favor opposite magnetizations. Thus, the Néel vector can be reversed by switching the net magnetization through uniaxial strain, under a constant magnetic field or by constructing heterostructure device with a ferromagnetic layer. Moreover, the tunable half-metallic state and valley polarization in 2D $VCrTe_2O$ offers new opportunity to realize novel exotic quantum phenomena such as field-free reversable fully spin-polarized currents, anomalous Hall effect, [4,6,7] spin Hall effect [12-14,24] and valley Hall effect [36,37,41], expanding the 2D magnetism landscape and offer pathways for energy-efficient spintronic applications.

As a matter of fact, $V_2Te_2O$ has been experimentally synthesized via a topochemical deintercalation of interlayer $Rb^+$ cations in $Rb_{1-\delta}V_2Te_2O$ [38], which exhibits metal transport property, consisting with our first-principles calculations. Since $V_2Te_2O$ belongs to van der Waals materials, 2D $V_2Te_2O$ sample can be obtained by exfoliate method. [2-9] Most recently, $RbCr_2Se_2O$ has also been experimentally compounded, [43] which indicates that $Rb_{1-\delta}VCrTe_2O$ might be synthesized by adjusting the Molar ratio of the raw materials and heating conditions, [38,43,44] suggesting a topochemical deintercalation method in synthesizing $VCrTe_2O$. Furthermore, AFiM can also be induced in other altermagnets by introducing the disparity of two sublattice spins connecting by the special symmetry, which deserves to be explored in future experiments.

## 4. Conclusion

In summary, we establish a novel class of two-dimensional collinear magnets, termed 2D alternating ferrimagnets, which are distinct from conventional antiferromagnets and the recently identified altermagnets. The systems exhibit concurrent alternating momentum-dependent band splitting and spin-dependent band shifting, as captured by a 2D tight-binding model. We further demonstrate two feasible strategies for experimental realization: strain engineering and elemental substitution based on prototypical 2D altermagnets. Through first-principles calculations and Monte Carlo simulations, the alternating ferrimagnetism is verified in monolayer $VCrTe_2O$ and strained $V_2Te_2O$, elucidating strain-tunable properties such as net magnetization, half-metallic behavior and valley polarization. This work underscores the potential of 2D alternating ferrimagnets to invigorate the development of advanced spintronic devices.

## 5. Methods

**Tight-binding model.** We transform the Hamiltonian (equation 1) into the reciprocal space and set $|\delta_{2x/y}| = 1$ to simplify the formula, then obtain the bands:

$$E^{up} = (t_a^x + t_b^x)\cos k_x + (t_a^y + t_b^y)\cos k_y - M_a - M_b \pm \sqrt{[(t_a^x - t_b^x)\cos k_x + (t_a^y - t_b^y)\cos k_y - \frac{\Delta}{2} - \frac{1}{2}M_a + \frac{1}{2}M_b]^2 + 4t^2(1+\cos k_x)(1+\cos k_y)} \quad (3)$$

$$E^{dn} = (t_a^x + t_b^x)\cos k_x + (t_a^y + t_b^y)\cos k_y + M_a + M_b \pm \sqrt{[(t_a^x - t_b^x)\cos k_x + (t_a^y - t_b^y)\cos k_y - \frac{\Delta}{2} + \frac{1}{2}M_a - \frac{1}{2}M_b]^2 + 4t^2(1+\cos k_x)(1+\cos k_y)} \quad (4)$$

The spin-splitting bands of FMs, AFMs, AMs, AFiMs can be obtained by setting these parameters and considering the symmetry.

**First-principles calculation.** The first-principles calculations are performed using the Vienna Ab initio Simulation Package. The exchange-correlation effects were treated with the generalized gradient approximation (GGA) in the Perdew-Burke-Ernzerhof (PBE) form. To account for the strong correlation of V/Cr 3d electrons, the GGA + *U* method with an effective Hubbard parameter $U_{eff}$ = 3.0/3.5 eV was employed. [24,45] The energy convergence threshold between consecutive electronic steps was set to $10^{-7}$ eV. A vacuum layer of 20 Å was introduced perpendicular to the monolayer to prevent interlayer interactions between periodic images. Furthermore, a plane-wave energy cutoff of 600 eV and a k-point grid of 16 × 16 × 1 were used for Brillouin zone sampling.

**Monte Carlo simulation.** The Monte Carlo (MC) simulations with the Metropolis algorithm are performed for studying the temperature-dependent magnetization and Néel vector of a 2D 100 × 100 square supercell for $V_2Te_2O$ and $VCrTe_2O$, based on the explicitly resolved spin Hamiltonian. The zero-field cooling approach is adopted, and $8\times10^5$ steps are discarded for equilibrium consideration and another $2\times10^5$ steps are retained for statistic averaging.

### Notes

The authors declare no competing financial interest.

## ACKNOWLEDGMENTS

This work was supported by the National Natural Science Foundation of China (Grants No. U22A20117, No. 12595330, No. 12574048, No. 12504123 and No. 52371243), the Guangdong Provincial Quantum Science strategic initiative (Grant No. GDZX2401002), the Guangdong Basic and Applied Basic Research Foundation (Grant No. 2024A1515012665), Jiangsu Funding Program for Excellent Postdoctoral Talent and the Postdoctoral Fellowship Program China of CPSF (Grant No. GZB20250766).The first-principles calculations in this work are supported by the Beijing Super Cloud Computing Center (BSCC).

## Figures

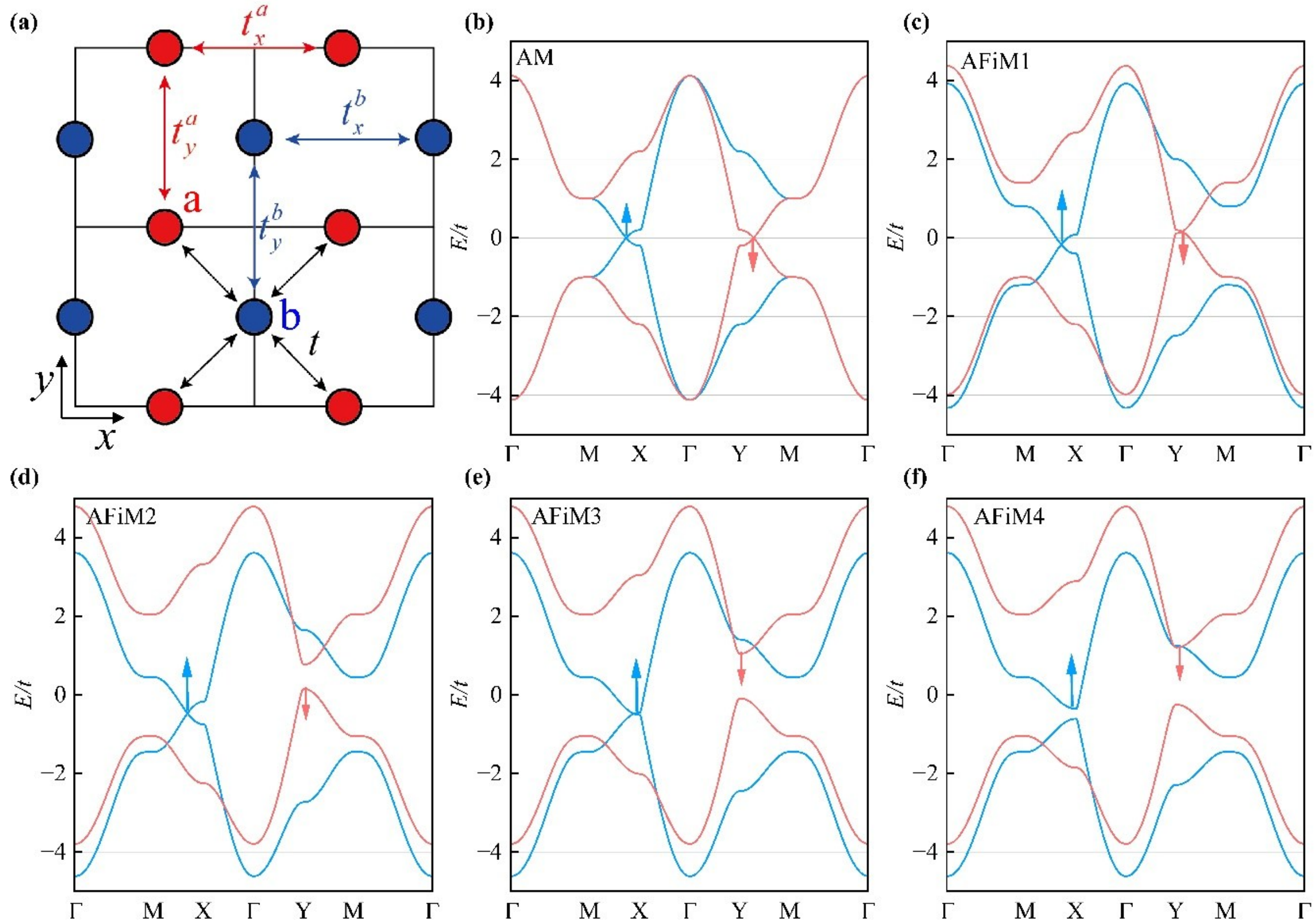

**Figure 1**. Energy band of the TB model for altermagnets and alternating ferrimagnets. (a) The illustration of the 2D model. (b) Energy band of altermagnets with alternating momentum-dependent spin splitting in condition that $t_x^a = t_y^b = 0.3t$, $t_y^a = t_x^b = -0.3t$, $\Delta = 0$, $-M_a = M_b = t$. Energy band of alternating ferrimagnets with concurrent alternating momentum-dependent spin splitting and spin-dependent band shifting, including AFiM1 (c) in condition that $t_x^a = 0.3t$, $t_y^b = 0.32t$, $t_y^a = -0.3t$, $t_x^b = -0.32t$, $\Delta = 0.2t$, $M_a = -t$, $M_b = 1.2t$, AFiM2 (d) in condition that $t_x^a = 0.3t$, $t_y^b = 0.32t$, $t_y^a = -0.3t$, $t_x^b = -0.32t$, $\Delta = 0.6t$, $M_a = -t$, $M_b = 1.5t$, AFiM3 (e) in condition that $t_x^a = 0.24t$, $t_y^b = 0.25t$, $t_y^a = -0.24t$, $t_x^b = -0.25t$, $\Delta = 0.6t$, $M_a = -t$, $M_b = 1.5t$, and AFiM4 (f) in condition that $t_x^a = 0.2t$, $t_y^b = 0.21t$, $t_y^a = -0.2t$, $t_x^b = -0.21t$, $\Delta = 0.6t$, $M_a = -t$, $M_b = 1.5t$.

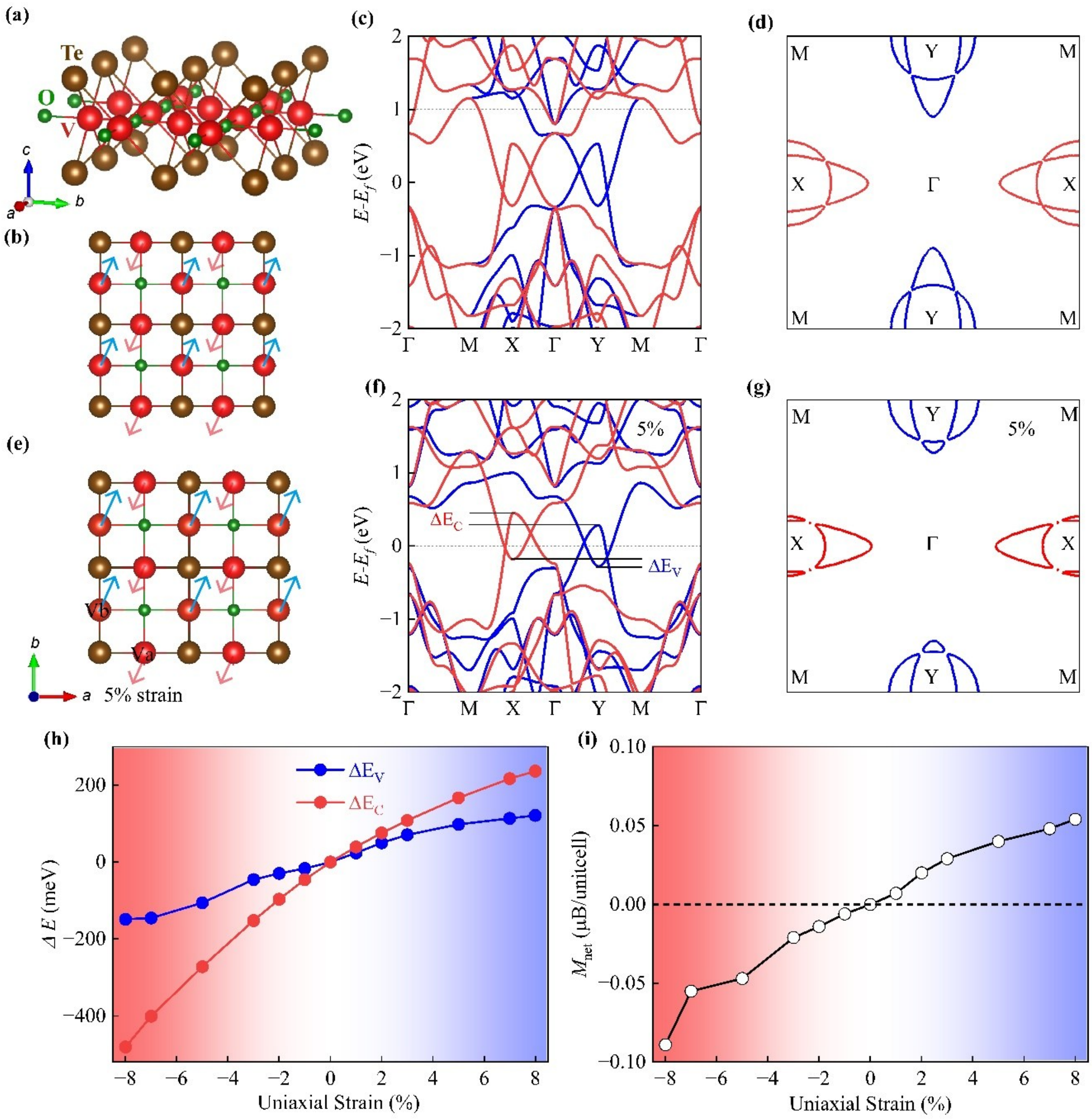

**Figure 2.** The side view of crystal structure of monolayer $V_2Te_2O$ (a). The top view of crystal structure and altermagnetic ground state in monolayer $V_2Te_2O$ (b) and its band structure (c), Fermi surface within the first Brillouin zone (d). The top view of monolayered $V_2Te_2O$ under 5% uniaxial tensile strain along *a* direction (e) and its band structure (f), Fermi surface (g). The blue lines in the band structure represent spin up while red lines denote bands for spin down. (h) Evolution of spin-dependent band shifting energy of valance and conduction band between bands located at X and Y near the Fermi level with the uniaxial train for monolayered $V_2Te_2O$. (i) Evolution of the net magnetization of monolayered $V_2Te_2O$ with uniaxial strains, demonstrating a strain-tunable net magnetization.

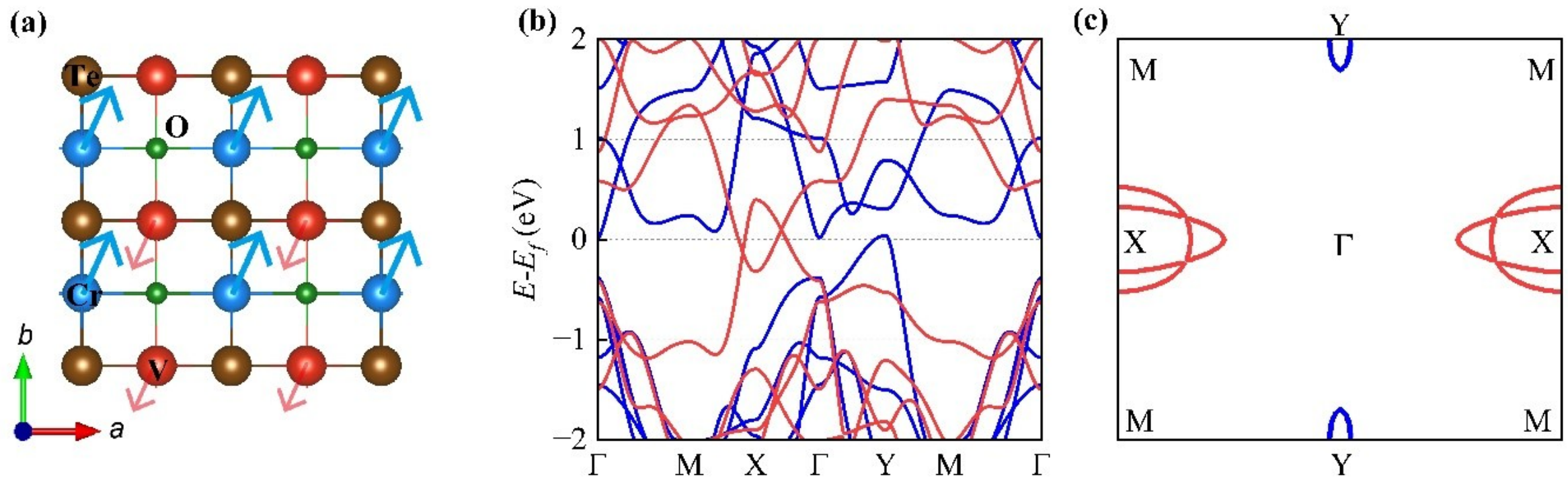

**Figure 3.** The top view of the structure of monolayered $VCrTe_2O$ (a) and its band structure (b), Fermi surface (c). The blue lines in the band structure represent spin up while red lines denote bands for spin down.

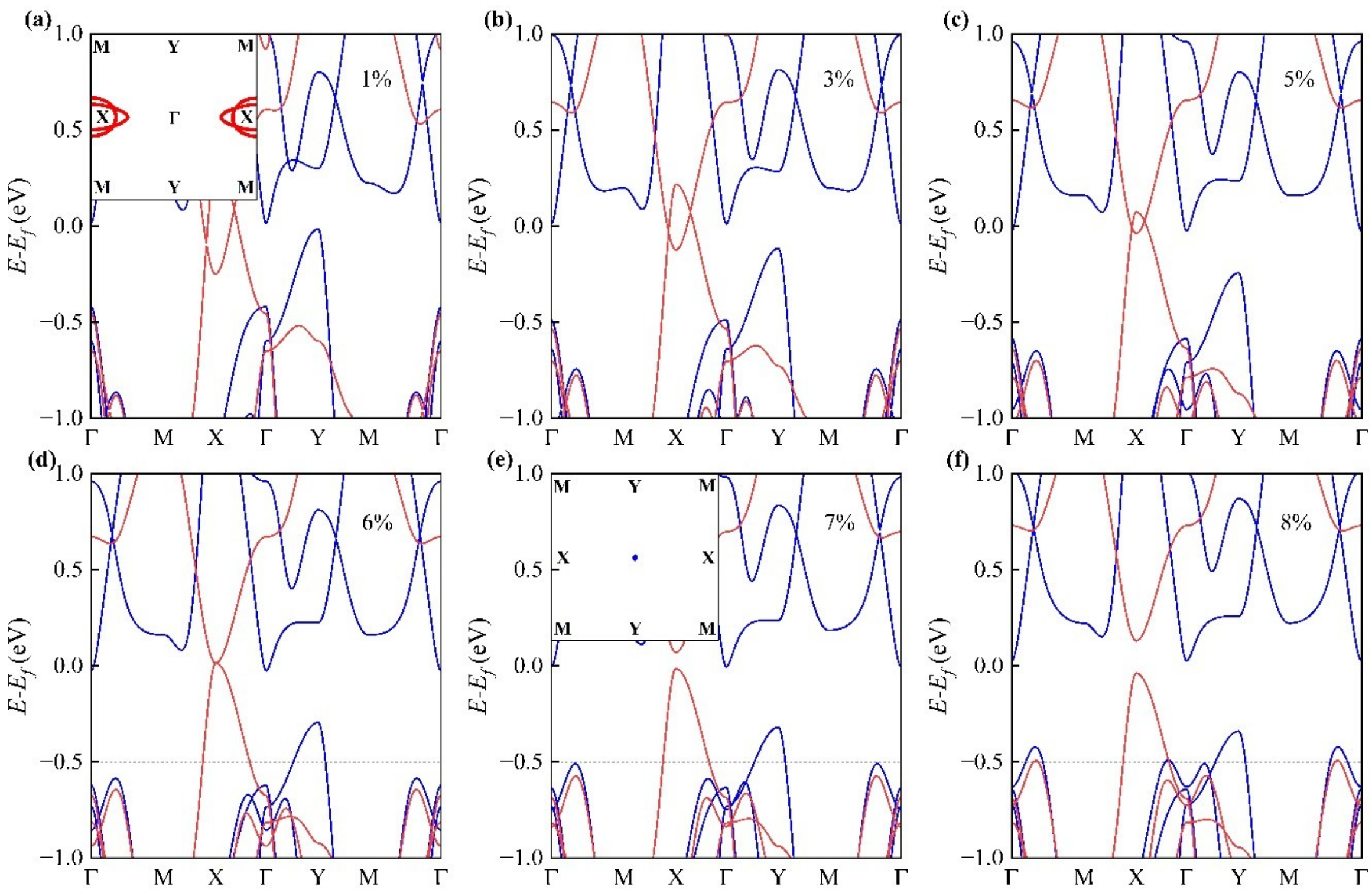

**Figure 4.** The evolution of band structure in monolayered $VCrTe_2O$ with biaxial tensile strains. Band structure under (a) 1% strain, emerging a half-metallic state with spin-down polarization at the Fermi level (inset shows Fermi surface), (b) 3% with half-metallic state, (c) 5% strain, (d) 6% strain with the nodal loop for spin down transforms into crossing point, (e) 7% strain (inset shows Fermi surface), emerging half-metallic state with opposite spin polarization (spin-up) at the Fermi level, (e) 8% tensile strain, entering semiconductor state. The valley polarization effect between X point and Y point is observed under the 7% and 8% strain. The blue lines in the band structure represent spin up while red lines denote bands for spin down.

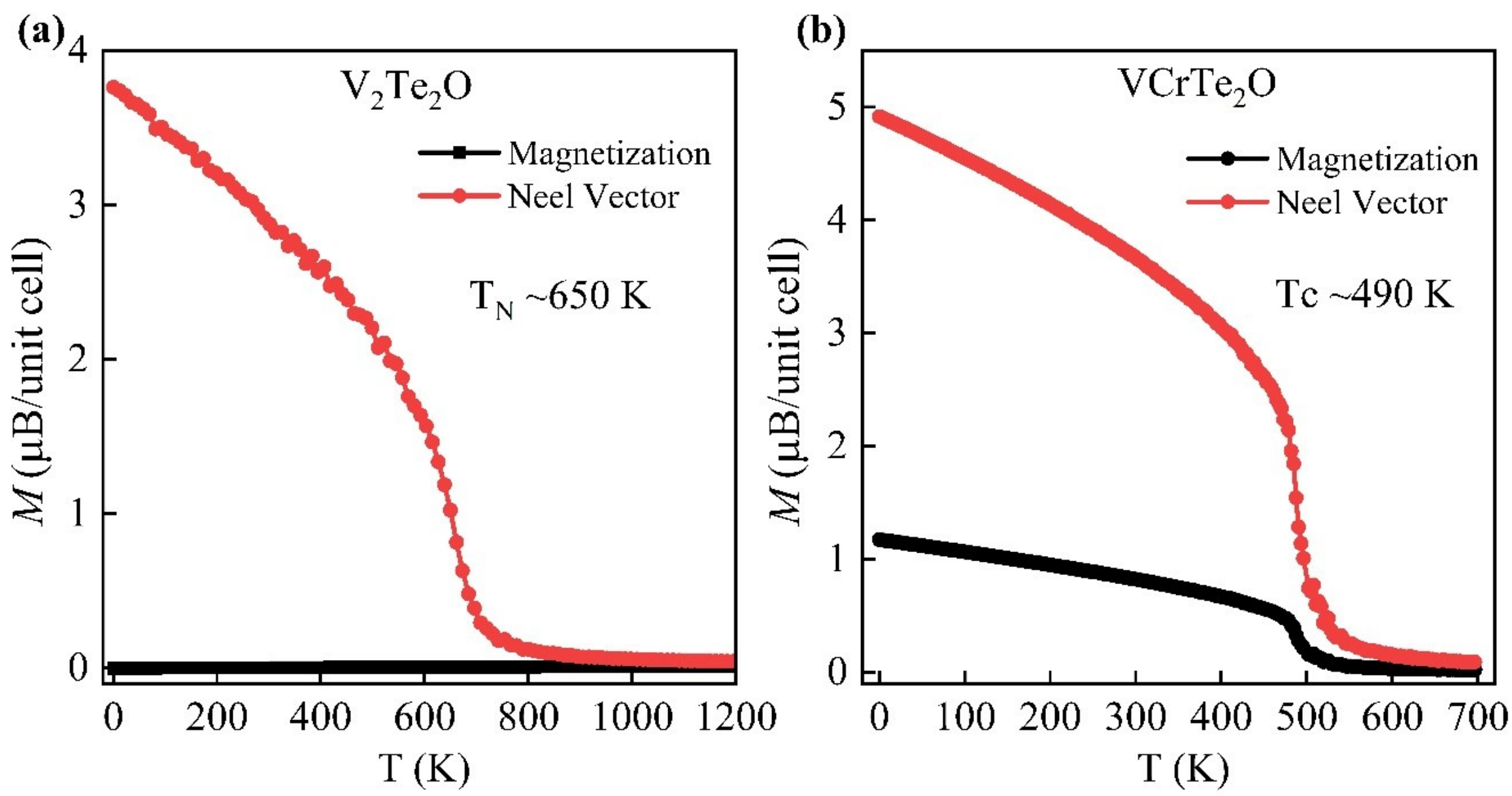

**Figure 5.** The temperature-dependent magnetization and Néel vector of monolayered $V_2Te_2O$ (a) and monolayered $VCrTe_2O$ (b), both exceed room temperature.

**Tables**

| | $a$ (A) | $b$ (A) | $M_a$ ($\mu_B$) | $M_b$ ($\mu_B$) |
|---|---|---|---|---|
| $V_2Te_2O$ | 3.93 | 3.93 | -1.811 | 1.811 |
| $V_2Te_2O$ under 5% strain | 4.13 | 3.93 | -1.861 | 1.874 |
| $VCrTe_2O$ | 4.01 | 3.88 | -1.873 | 3.046 |

**Table 1.** Lattice constant and magnetic moment for monolayered $V_2Te_2O$, $V_2Te_2O$ under 5% uniaxial strain, and $VCrTe_2O$.

| | *FM* | *AF1* | *AF2* | *AF3* | *AF4* | *AF5* | *AF6* |
|---|---|---|---|---|---|---|---|
| $V_2Te_2O$ | -116.85 | -118.20 | -116.65 | -116.54 | -116.72 | -116.64 | -116.80 |
| $VCrTe_2O$ | -116.06 | -117.80 | -116.81 | -116.70 | -116.72 | -116.66 | -116.67 |

**Table 2.** Total energies of typical spin configurations (Figure S2) for 4×4 supercell in monolayered $V_2Te_2O$ and $VCrTe_2O$. The unit is meV.

| | $J_1$ | $J_{2x_vv}$ | $J_{2y_vv}$ | $J_{2x_cc}$ | $J_{2x_cc}$ | $K_x$ | $K_y$ | $K_z$ |
|---|---|---|---|---|---|---|---|---|
| $V_2Te_2O$ | 26.26 | 3.61 | -2.66 | | | *0* | *0* | 0.07 |
| $VCrTe_2O$ | 12.73 | -2.21 | -0.11 | 1.46 | 1.18 | -0.14 | 0.22 | 0 |

**Table 3.** The Heisenberg parameters for monolayered $V_2Te_2O$ and $VCrTe_2O$. The positive and negative signs of $J$ indicate the antiferromagnetic and ferromagnetic exchange coupling. The unit is meV.